\documentclass{mn2e}
\usepackage{epsfig}
\def\msun{{\rm M_{\odot}}}

\def\me{{\dot M_{\rm Edd}}}

\def\mo{{\dot M_{\rm out}}}
\def\le{{L_{\rm Edd}}}

\def\ltsima{$\; \buildrel < \over \sim \;$}
\def\simlt{\lower.5ex\hbox{\ltsima}}
\def\gtsima{$\; \buildrel > \over \sim \;$}
\def\simgt{\lower.5ex\hbox{\gtsima}}
\def\sgra{Sgr~A$^*$}

\newcommand\ledd{{L}_{\rm Edd}}

\newcommand\mbh{{\,{\rm M}_{\rm bh}}}

\newcommand{\apj}{ApJ}

\newcommand{\mnras}{MNRAS}
\newcommand{\aap}{A\&A}
\newcommand{\araa}{ARA\&A}
\newcommand{\apjl}{ApJL}

\newcommand{\nat}{Nature}

\def\del#1{{}}

\title[Fermi Bubbles: Echoes of the Last Quasar Outburst?] {The Milky Way's Fermi Bubbles: Echoes of the Last Quasar Outburst?}

\author[K. Zubovas, A.R. King, S. Nayakshin] {K. Zubovas$^{1}$,
  A.R. King,$^{1}$ S. Nayakshin$^{1}$ \\
$^1$Theoretical Astrophysics
  Group, University of Leicester, Leicester LE1 7RH}
      
\date{\today}

\volume{000}

\pagerange{\pageref{firstpage}--\pageref{lastpage}} \pubyear{2005}

\begin{document}

\label{firstpage}

\maketitle

\begin{abstract}

{\it Fermi}--LAT has recently detected two gamma ray bubbles disposed
symmetrically with respect to the Galactic plane. The bubbles have
  been suggested to be in a quasi--steady state, inflated by ongoing star
  formation over the age of the Galaxy. Here we propose an alternative picture
  where the bubbles are the remnants of a large--scale wide--angle outflow
  from \sgra, the SMBH of our Galaxy. Such an outflow would be a
natural consequence of a short but bright accretion event on to \sgra\ if
  it happened concurrently with the well known star formation event in the
inner 0.5 pc of the Milky Way $\sim 6$~Myr ago. We find that the
  hypothesised near--spherical outflow is focussed into a pair of symmetrical
lobes by the greater gas pressure along the Galactic plane. The outflow
shocks against the interstellar gas in the Galaxy bulge. Gamma--ray emission
could be powered by cosmic rays created by either \sgra\ directly or
  accelerated in the shocks with the external medium.  The Galaxy disc
remains unaffected, agreeing with recent observational evidence that
supermassive black holes do not correlate with galaxy disc properties. We
estimate that an accreted mass $\sim 2 \times 10^3\msun$ is needed for the
accretion event to power the observed {\it Fermi}--LAT lobes. Within a
factor of a few this agrees with the mass of the young stars born during the
star formation event. This estimate suggests that roughly $50$\% of the gas
was turned into stars, while the rest accreted onto \sgra. One interpretation
of this is a reduced star formation efficiency inside the
  \sgra\ accretion disc due to stellar feedback, and the other a peculiar
mass deposition geometry that resulted in a significant amount of gas falling
directly inside the inner $\sim 0.03$ pc of the Galaxy.


\end{abstract}

\begin{keywords}{galaxies:evolution - galaxies: Milky Way -
    quasars:general - 
black hole physics - accretion}
\end{keywords}

\section{Introduction}

\subsection{\sgra\ -- the SMBH of the Milky Way}

Sgr A$^*$ is the supermassive black hole (SMBH) in the nucleus of our
Galaxy. Its mass $\mbh \simeq 4 \times 10^6 \msun$ (Sch{\"o}del et al. 2002,
Ghez et al. 2005, 2008) makes it directly comparable with SMBH in other
galaxies. The Soltan relation (Soltan 1982) implies that most of the mass of
these black holes was gained through luminous accretion.  Yet by comparison
with active galactic nuclei (AGN) \sgra\ is famously dim. It is spectacularly
faint both in X-rays (less than $\sim 10^{-11} \ledd$, where $\ledd \sim$ a
few $\times 10^{44}$ erg s$^{-1}$ is its Eddington luminosity; Baganoff et
al. 2003) and in the near infrared (Genzel et al. 2003), prompting suggestions
of a radiatively inefficient accretion flow (Narayan 2002, and references
therein). Currently, \sgra\ appears to be fed by accretion of gas captured
(Cuadra et al. 2006) from the winds of the young massive stars populating the
inner $\sim 0.5$ pc of the Galaxy (Paumard et al. 2006). However, X--ray
reflection nebulae suggest that \sgra\ might have been much brighter a few
hundred years ago, with luminosity of a few $\times 10^{39}$ erg s$^{-1}$
(e.g., Revnivtsev et al 2004, Ponti et al 2010). This may reflect variations
in the wind feeding rate of \sgra\ caused by changes in the stellar orbits of
the most important wind--producing massive stars (Cuadra et al. 2008), or
longer time scale feeding events from a few pc-scale molecular gas reservoirs
(Morris et al. 1999).

\sgra\ is also famous as the site of a recent ($\sim 6 \times 10^6$ yr ago)
star formation event in one and perhaps two stellar discs (Levin \& Beloborodov 2003,
Genzel et al. 2003, Lu et al. 2009) on scales of $\sim 0.03 - 0.5$ pc from the
SMBH. The observed (e.g. Paumard et al. 2006) and theoretically constrained
(Nayakshin et al., 2006) mass of the young stars is around a few times $10^3
\msun$, perhaps even $10^4 \msun$. Significantly, there is currently no trace 
of even a remnant gaseous
disc near \sgra\ (Cuadra et al. 2003, Paumard et al. 2004). This led Nayakshin
\& Cuadra (2005) to question whether \sgra\ failed to become a quasar because
this recent star formation event consumed nearly all the available gas in the
central parsec of the Milky Way. They noted that this could be constrained
with future observations: ``a past bright AGN phase should also leave a hot
buoyant radio bubble in the Milky Way halo''.

\subsection{The Fermi--LAT gamma--ray lobes}

The recent {\it Fermi}--LAT observations by Su et al. (2010) show that the
Milky Way has a pair of gamma--ray lobes, symmetrical about its dynamical
centre (\sgra) and about the Galactic plane. The lobes extend $\sim
  5$~kpc from the plane, but have a narrow ($d \sim 100$ pc) waist
along it. The limbs of the lobes coincide with the extended structure
seen in medium--energy X--rays by ROSAT (Snowden et al. 1997).  The lobes have
gamma--ray luminosity $L_{\gamma} \simeq 4\times 10^{37}~{\rm erg\, s}^{-1}$,
and their total energy content is at least $\sim 10^{54 - 55}$~erg.

Su et al. (2010) considered numerous physical processes that could give
  rise to the bubble structure and provided a constraint that if they are
  older than a few $\times 10^6$ yr, the gamma-ray emission must be powered by
  ions rather than electrons due to a short cooling time of the
  latter. Crocker et al. (2011) and Crocker \& Aharonian (2011) detailed these
  arguments further and suggested that the emission is powered by Cosmic Ray
  (CR) protons rather than electrons. They further consider a quasi--steady
  state model in which the CR protons are continuously injected by supernova
  explosions. CR protons and heavier ions are then trapped inside the bubbles
  for approximately the age of the Galaxy.

Alternatively, the {\it Fermi}--LAT lobes could be a more recent
  feature. For example, Mertsch \& Sarkar (2011) argue that the spectral and
  morphological details of the emission are incompatible with hadronic
  radiation, leaving electrons as the main energy source. In that case, the
  mechanism of inflating the bubbles is then unlikely to be of
 star
  formation origin. One would require $\sim 10^5$ recent Type II supernovae to
  provide the energy content of the bubbles, which is far higher than can be
  realistically expected from the inner $\sim 100$ pc. Cheng et al. (2011)
  thus argued that the bubbles are inflated by episodic \sgra\ activity
  caused
 by tidal disruptions of stars passing too close to \sgra. Guo \&
  Mathews
 (2011) performed hydrodynamical numerical simulations of jets
  launched by
 \sgra\ and showed that the {\it Fermi}--LAT observations are
  qualitatively
 consistent with their simulations if the jets were launched
  $\sim 1-2$ Myrs
 ago.

\subsection{\sgra\ feedback: when and how?}

In this paper we shall argue that \sgra\ is a very natural candidate for
  the source of the energy that inflated the gamma-ray lobes.  As noted above, 
the Galactic Centre underwent a peculiar star formation event  
localised to the inner 0.03 -- 0.5 pc about 6~Myr ago (Paumard et al.  2006).
 Thus, a plausible scenario is that not all of the gas deposited into the 
central pc of the Milky Way went into making the young stars, and a fraction 
of it was accreted by \sgra, as found in realistic simulations of the process 
(e.g., Bonnell \& Rice 2008, Hobbs \& Nayakshin 2009). Thus \sgra\ is 
likely to have had a short but very bright quasar phase concurrent with the 
star formation event $\sim 6$ million years ago.

We further argue that the observed highly symmetrical lobes are
  unlikely to have originated from a jet outflow. To obtain the qualitative
  agreement with the observed shape of the lobes, Guo \& Mathews (2011)
  directed their jets perpendicular to the plane of the Galaxy. We believe
  this would be unlikely. Radio surveys show that jet directions are
completely uncorrelated with the large--scale structure of the host galaxies
(Kinney, 2000; Nagar \& Wilson, 1999). Furthemore, the observed orientations
of the stellar discs in the central pc of the Galaxy (see Paumard et al.
 2006) are inclined at very large
angles to the Galactic plane. The jets are likely to be fed by gas discs
  oriented similarly to the stellar discs. We would therefore expect that
accretion of gas onto \sgra\ $\sim 6$ million years ago would result in jets
directed at very large angles to the symmetry axis of the lobes, contradicting
observations.

In contrast, a symmetrical pair of lobes with a narrow waist along the galaxy
plane is natural if an isotropic outflow from near the black hole encounters
higher gas densities along this plane than perpendicular to
it. Near--spherical outflows like this are a direct consequence of
super--Eddington disc accretion (Shakura \& Sunyaev 1973; King \& Pounds
2003) and offer a plausible explanation for the $M$--$\sigma$ relation (Silk \&
Rees 1998; King 2003, 2005). 

The paper is structured as follows. We first discuss the simpler and better
understood quasi-spherical AGN outflows in \S 2, and then we consider the more
complicated case of the present day Milky Way nucleus in \S 3. The
implications of the quasar outburst for the observed gamma-ray lobes are
elucidated in \S 4, while \S 5 spells out ramifications for the poorly
understood problem of star formation versus gas accretion in the central
parsecs of AGN. We note that our approach here is to try to reproduce the
  main energetics of the lobes and their morphology rather than to produce
  detailed spectral models. We assume that \sgra\ outflow either carries with
  it CR protons created near the black hole, or that the CR protons are
  accelerated on shock fronts where the outflow runs into the interstellar
  medium.

\section{Spherical outflows}

In regions close to the black hole, the AGN outflows are revealed through
blueshifted absorption lines in X--ray emission (Pounds et al. 2003a, b;
King 2010a). Tombesi et al. (2010a, b) show that they are present in more
than 35 percent of a sample of over 50 local AGN, and deduce that their solid
angles are large (certainly $> 0.6 \times 2\pi$, and probably greater). The
observed absorption columns imply that in many cases the outflows are quite
recent (few years), suggesting that outflows are an almost ubiquitous feature
of central black hole activity (King 2010b).

Although supermassive black holes (SMBHs) in galaxy centres frequently accrete
at the Eddington rate, accretion at significantly higher rates requires
extreme conditions (cf King, 2010a). Accordingly we consider cases where the
accretion rate far from the SMBH only mildly exceeds $\me$, and both the
central accretion rate and the outflow rate $\mo$ are $\simeq \me$. Then the
outflow has scattering optical depth $\sim 1$, and the photons driving it
typically scatter only once before escaping. The front--back symmetry of
electron scattering means that the outflow momentum must be of the same order
as the original photon momentum, i.e.
\begin{equation}
\mo v \simeq {\le\over c}
\end{equation}
so that the outflow velocity $v \sim \eta c$, where $\eta \sim 0.1$ is
the accretion efficiency (e.g. King \& Pounds 2003; King 2010a). The
wind flows with essentially constant velocity $v$ until it shocks
against the interstellar gas of the host galaxy, driving a second
shock outwards into this ambient medium and sweeping its up into a
shell. A simple representation of the interstellar density is the
isothermal distribution
\begin{equation}
\rho(R) = {f_g\sigma^2\over 2\pi G R^2}
\label{rho}
\end{equation}
where $f_g$ is the gas fraction, and $\sigma$ is the velocity dispersion. The
average cosmological value of $f_g$ is $f_c\simeq 0.16$. 

Within this model, then, in galaxies with large $\sigma \simgt 150$ km
s$^{-1}$, Eddington outflows tend to sweep the
 vicinity of the hole clear of
gas of density (\ref{rho}) and prevent
 further accretion and growth,
establishing the $M - \sigma$ relation
 for the black hole mass (King 2003;
2005). At smaller values of
 $\sigma$, any effect of this kind is outdone by
the effects of mass
 loss from nuclear star clusters. These sweep out the gas
(McLaughlin
 et al., 2006; Nayakshin et al., 2009) and establish an offset
$M$--
 $\sigma$ relation between the total cluster mass and the bulge
velocity
 dispersion (Ferrarese et al. 2006, Fig. 2, middle panel). The Milky
Way is probably a member of this star--cluster
 dominated class of galaxies,
and indeed its SMBH mass lies
 significantly below the value predicted from
the $M$--$\sigma$ relation (see, e.g. Greene et al. 2010, Fig. 9).

The double shock pattern caused by the impact of an Eddington outflow
on the host interstellar medium must move radially outwards from the
vicinity of the black hole. The nature of this motion depends
crucially on whether or not the shocked wind cools within the flow
time. If cooling is effective, most of the energy injection rate
\begin{equation}
\dot E = {1\over 2}\mo v^2 = {\eta^2c^2\over 2}\mo = {\eta\over
  2}\le
\end{equation}
is lost to radiation, and only the ram pressure of the outflow is
communicated to the host ISM. This is a momentum--driven flow. If
instead the flow does not cool, the shocked wind expands
adiabatically, doing $P{\rm d}V$ work against the swept--up
interstellar medium. This is an energy--driven flow, which expands
faster through the ISM than a momentum--driven one. 

\section{{\it Fermi}--LAT lobes as quasi-spherical outflows}

The gamma--ray lobes observed by the {\it Fermi}--LAT instrument are very wide
features that we shall first consider approximately quasi-spherical. For the
present day Milky Way and directions well out of the Galactic plane, we expect
$f_g$ to be significantly less than $f_c$, so we parametrize $f_g$ as $f_g =
1.6 \cdot 10^{-3} f_{0.01}$, where $f_{0.01} \sim 1$ is a dimensionless free
parameter of the model.

We now ask if the shocked gas cools in conditions appropriate for the
outburst.  The outflow speed $v \simeq 0.1c$ implies a shock temperature $T_s
= (3m_p/16k)\eta^2c^2 \simeq 1.6 \times 10^{10}$~K. This is much higher than
the Compton temperature $\sim 10^7$~K of the SMBH accretion flow, so when the
shock is sufficiently close to the hole, Compton cooling by the central
radiation field is very effective and enforces momentum--driven flow. As the
shock reaches a critical radius $R_{\rm en}$ the radiation field becomes too
dilute to cool it. Also, the shocked wind has far too low a density to cool
effectively by atomic or free--free processes, so the flow becomes
energy--driven (King 2003; King et al. 2011).  For the parameters of Sgr A*
(mass $M_{\rm BH} \simeq 4\times 10^6\msun$, velocity dispersion $\sigma
\simeq 100~{\rm km\, s^{-1}})$ the transition to energy--driven flow occurs at
a radius
\begin{equation}
R_{\rm en} \simeq 15f_{0.01}^{1/2}~{\rm pc}
\label{rcool}
\end{equation}
(cf eqs 8 - 10 of King 2003). 

Even at the cosmological gas fraction ($f_{0.01} = 100$) the estimate
(\ref{rcool}) is so small compared with the size of the gamma--ray
lobes that we can regard the outflow as essentially always
energy--driven in directions away from the Galactic plane. In an
energy--driven outflow, the shocked wind density driving the expansion
is always much lower than the density of the swept--up interstellar
medium outside it. This makes the shock interface inherently
Rayleigh--Taylor unstable (cf King 2010b). The hot shocked gas mixes
with cool dense interstellar gas throughout the flow in directions
away from the Galactic plane. This mixture is clearly a promising site
for gamma--ray emission. Within the Galactic plane the gas density is
far higher, and we expect little expansion  (see also \S 4). This kind of outflow thus
naturally produces the main qualitative features of the {\it Fermi}--LAT
gamma--ray map: extensive gamma--ray emitting lobes placed
symmetrically on each side of the Galactic plane, with a narrow waist
in the plane.


Energy--driven outflows rapidly attain a constant speed
\begin{equation}
v_e = \left[\frac{2 \eta \sigma^2 c}{3 b}\frac{f_c}{f_g}\frac{M}{M_{\sigma}}\right]^{1/3}
 \simeq 1640 \; \sigma_{100}^{2/3} f_{0.01}^{-1/3} \; \mathrm{km} \; \mathrm{s}^{-1}
\label{en}
\end{equation}
in the bulge of a galaxy (King 2005). Here the factor $b \la 1$ allows
for some collimation of the outflow, $M_{\sigma}$ is the predicted value
of the SMBH mass in the Milky Way from the $M - \sigma$ relation and
$M \simeq 0.2 M_\sigma$ is the mass of \sgra.

For the rest of this Section we model the outflow away from the disc
plane as a sector of a spherical flow. If the Eddington
accretion phase lasts for a time $t_{\rm acc}$, the shock reaches
radius
\begin{equation}
R_0 \simeq v_et_{\rm acc}
\label{R0}
\end{equation}
when the quasar phase ends. However the shocked wind gas is able to
drive further expansion, which finally stalls at a radius
\begin{equation}
R_{\rm stall} \simeq {v_e\over \sigma}R_0 \simeq  {v_e^2\over
  \sigma}t_{\rm acc}
\label{rstall}
\end{equation}
after a time 
\begin{equation}
t_{\rm stall} \simeq 0.5\biggl({v_e\over\sigma}\biggr)^2t_{\rm acc}
\label{tstall}
\end{equation}
(King et al. 2011)
so that
\begin{equation}
t_{\rm stall} = {R_{\rm stall}\over 2\sigma}
\label{stall}
\end{equation}

We now apply these considerations to our Galaxy, and in particular the
gamma--ray lobes.
If the outflow producing the observed lobes had stalled, we
would have $R_{\rm stall} \sim 5$~kpc, which from (\ref{stall})
requires $t_{\rm stall} = 15$~Myr. This would mean that the outflow
was produced well before the last accretion event in the Galactic
Centre, which appears unlikely.

If instead we assume that the gamma--ray lobes were produced in this
event, we must conclude that the energy--driven outflow is still
proceeding, with a mean velocity $\left<v\right> \simeq 1000~{\rm km\, 
s^{-1}}$ over its lifetime. This
is lower than the shell velocity during the quasar phase, which might
have been as large as $v_e \simeq 1600$ km s$^{-1}$ (cf. the figures in
King et al. 2011). This is compatible with (\ref{en}) if $f_{0.01} \sim
1$, i.e. if the gas fraction in the lobes is about $1\%$ of the
cosmological value. Requiring $t_{\rm stall} > 6$~Myr in
(\ref{tstall}) now gives $t_{\rm acc} \ga 5\times 10^4$~yr. At the Eddington
accretion rate $\simeq 4\times 10^{-2}\;\msun\, {\rm yr}^{-1}$
appropriate for the mass of Sgr A* this gives the total mass accreted
by the black hole during the quasar phase as $\Delta M \ga
2 \times10^3\;\msun$. This is comparable to the total expected if the hole
accreted the disc mass
\begin{equation}
M_{\rm disc} \sim {H\over R} M_{\rm BH} \simeq 8000\;\msun
\end{equation}
within the self--gravity radius where the ring of young stars formed
(cf eqs 7, 12 of King \& Pringle, 2007). This estimate is also consistent with
the results of Nayakshin \& Cuadra (2005).

In this picture the mass of wind expelled from the vicinity of the hole must
be $\sim\Delta M$. Almost all of its kinetic energy is retained by the
outflow.  This energy is of order
\begin{equation}
E_{\rm lobes} \sim \frac{\eta^2}{2} \Delta M c^2 \ga 2 \times10^{55}~{\rm erg},
\end{equation}
somewhat above the minimum required by observation. At the current
gamma--ray luminosity the lifetime is $\sim 10^{10}$~yr, but there may
be other losses of course. We conclude that
the properties of the gamma--ray lobes are consistent with their
production in a short phase of Eddington accretion about 6~Myr ago.

\section{The narrow waist of the gamma-ray lobes}

So far we considered the outflow in a spherical geometry. However, it is well
known that the central $\sim 200$ pc of the Galaxy host as much as $\sim 10$\%
of all molecular gas of the Galaxy in a flattened, presumably disc-like,
configuration (Morris \& Serabyn 1996). We shall now show that this feature,
called the Central Molecular Zone (CMZ), could not have been significantly
affected by the hypothesised outflow from \sgra.

The mass of the molecular gas in the zone is $M_{\rm cmz} \sim 5\times
10^7\msun$. Its weight is 
\begin{equation}
W_{\rm cmz} \sim {G M_{\rm enc}(R_{\rm cmz}) M_{\rm cmz} \over R_{\rm cmz}^2}
= {2 M_{\rm cmz} \sigma^2 \over R_{\rm cmz}}\;,
\end{equation}
where $M_{\rm enc}(R_{\rm cmz})$ is the mass enclosed within radius $R_{\rm
  cmz}\sim 200$ pc. The outward force (momentum flux of the outflow incident
on the CMZ) in the isotropic outflow model is 
\begin{equation}
F_{\rm out} \sim {H \over R} {\ledd \over c}\;,
\label{fout}
\end{equation}
where $H/R \sim 0.2-0.3$ is the geometrical aspect ratio of the disc (see
Fig. 1 in Morris \& Serabyn 1996).

Comparing the two for the fiducial parameters accepted above, we have
\begin{equation}
{F_{\rm out} \over W_{\rm cmz}} \sim 0.1\;,
\label{no_chance}
\end{equation}
which shows convincingly that the outflow from \sgra, even in its full
``quasar'' mode, is not strong enough to disperse the CMZ since the latter is
simply too massive. This conclusion is reinforced by the fact that there is
also atomic and ionised gas in the region of the CMZ disc that would increase
$W_{\rm cmz}$ further.  

Another way to come to the same conclusion is through estimation of the gas
density in the midplane of the CMZ, for which we infer $\rho_{\rm cmz} \sim
5\times 10^{-22}$ g cm$^{-3}$ with the parameters mentioned above, whereas the
density of gas which could be driven away by a SMBH outflow, for a SMBH
obeying the $M-\sigma$ relation, is given by equation \ref{rho}, and is $\sim
10^{-22}$ g cm$^{-3}$ at $R=200$ pc and $\sigma = 100$ km s$^{-1}$.

We therefore conclude that the outflow along equatorial directions stalls. As
argued above, the outflow should then thermalise and expand away from the
plane of the symmetry, i.e., the Galactic plane. This would naturally explain
the two--lobe structure of the {\it Fermi}--LAT bubble.

We also note that a relatively geometrically thin distribution of the
molecular gas along the Galaxy plane in the CMZ justifies our assumption that
gas mass and density in the direction significantly away from the plane is
low. Indeed, the density at height $z \sim R$ away from the midplane is $\sim
\exp[-(R/H)^2]$ that in the midplane. Even at $H/R = 0.3$ this factor is about
  $10^{-3}$. We therefore estimate that the outflow should have an opening
  angle larger than $45^\circ$, and realistically in the range of $\sim
  60-70^\circ$.

Thus although outflows ultimately control black--hole growth, and materially
affect the Galaxy bulge, as shown by the $M - \sigma$ relation, they cannot
disperse the Galaxy disc. This fits very naturally with the recent conclusion
by Kormendy, Bender \& Cornell (2011) that SMBHs do not correlate
observationally with host galaxy discs.

\section{Accretion versus star formation}

One of the major puzzles in how SMBHs gain mass is the role of accretion disc
self--gravity in the outer $\simgt 0.03-0.1$ pc regions. Here we expect the
disc to be very cold and massive, provoking the conversion of gas into stars
(e.g., Paczynski 1978, Kolykhalov \& Sunyaev 1980, Goodman
2003). Hydrodynamical simulations of planar accretion discs in the regime
appropriate for the \sgra\ star formation episode lead to rapid gas depletion
through star formation (Nayakshin et al. 2007), leaving almost no fuel for the
SMBH.  The {\it Fermi}--LAT observations suggest that at least in the last
star formation episode near \sgra\, it managed to gain roughly the
same amount of gas as was used to make stars, which appears to be the first
observational evidence of this kind.

The interpretation of the physical significance of an amicable $\sim$ 50\%
split of the gas consumption between the stellar disc and \sgra\ is model
dependent. On the one hand, it may be an evidence that feedback from stars
inside the accretion disc is able to stave off star formation for long enough
to channel a sufficient amount of fuel to the SMBH (Thompson et al
2005). Alternatively, the dynamically hot structure of the young stars in the
central parsec of the Milky Way is best explained by a non--planar gas
deposition event resulting from, e.g., collisions of two massive gas clouds
(Hobbs \& Nayakshin 2009), collision between a cloud and the circumnuclear
disc, or capture of a large, turbulent giant molecular cloud (Wardle \&
Yusef--Zadeh 2008, Bonnell \& Rice 2008). If this is so, a fraction of the gas
may have small enough angular momentum to orbit within the innermost {\em non
  self--gravitating} disc region, avoiding the star formation catastrophe
entirely (King \& Pringle 2007, Nayakshin \& King 2007, Hobbs et al. 2010). In
particular, in all of the cases simulated by Hobbs \& Nayakshin (2009), the
mass of the gas captured inside their inner boundary (a non self--gravitating
part of the disc) was comparable with the mass required to fuel the {\it
  Fermi}--LAT lobes.

\section{Discussion}

We have shown that the shape and energy content of the gamma--ray
lobes observed by {\it Fermi} are consistent with the effects of a
near--isotropic outflow driven by the Milky Way's last Eddington
outburst, which we hypothesize to have been coincident with the well--known
star formation event in the central $0.5$ pc of the Galaxy
about 6~Myr ago.  The shape follows because such outflows
cannot penetrate the galaxy disc, in agreement with the recent conclusion
that SMBHs do not correlate observationally with galaxy discs
(Kormendy, Bender \& Cornell 2011). The accreted mass driving this event is
$\ga 2 \times10^3\msun$, comparable with the mass of the ring of stars born
in the event. This suggests that a significant fraction of gas captured into orbit
by the SMBH has low enough angular momentum to accrete rather than being
turned into stars. Because the outflow is Rayleigh--Taylor
unstable, the gamma--ray producing gas is homogeneously mixed though the
lobes.

Our model is similar to that of Crocker \& Aharonian (2011) in terms of
  assuming the CR protons (rather than electrons) produce the observed
  gamma-ray emission. We differ in the source and the age of the lobes. It is
  worth noting that if the bubbles are indeed very old quasi-static features
  as suggested by Crocker \& Aharonian (2011), then our estimates on the
  recent \sgra\ activity should be taken as a non-trivial upper
  limit for that activity.

The picture of the recent \sgra\ outburst developed here fits with a
general view that all galaxies are active and reach the Eddington limit from
time to time. This latter property is seen most obviously among galaxies
classified on other grounds as active, particularly in the very high fraction
of local AGN with high--speed outflows (Tombesi et al. 2010a, b).  It seems
likely that other `normal' galaxies may produce similar phenomena. However the
comparative dimness ($L_{\gamma} \sim 10^{37}~{\rm erg\, s^{-1}}$) of the
lobes may make these difficult to detect. A prediction of the present paper is
that the outermost parts of the lobes should still be expanding at $\sim
1000~{\rm km\, s^{-1}}$.

\section{Acknowledgments}

We thank the anonymous referee for extensive comments which allowed us to
significantly improve the clarity of our argument. We further thank Roland
Crocker and Mark Morris for illuminating discussions on the subject.

KZ is supported by an STFC studentship. Theoretical astrophysics research in Leicester 
is supported by an STFC Rolling Grant.

{}


\begin{thebibliography}{}

\bibitem[Baganoff et al.(2003)]{2003ApJ...591..891B} Baganoff, F.~K., et 
al.\ 2003, \apj, 591, 891 

\bibitem[Balick 
\& Brown(1974)]{1974ApJ...194..265B} Balick, B., \& Brown, R.~L.\ 1974, \apj, 194, 265 



\bibitem[Bonnell 
\& Rice(2008)]{2008Sci...321.1060B} Bonnell, I.~A., \& Rice, W.~K.~M.\ 2008, Science, 321, 1060 

\bibitem[Cheng et al.(2011)]{2011ApJ...731L..17C} Cheng, K.~S., Chernyshov, D.~O., Dogiel, V.~A., Ko, C.~M., Ip, W.~H.\ 2011, \apj, 731, L17 


\bibitem[Crocker \& Aharonian (2011)]{2011PhRvL.106j1102C} Crocker, R.~M., \& Aharonian, F.\ 2011, Physical Review Letters, 106, 1102 

\bibitem[Crocker et al. (2011)]{2011MNRAS.tmp..313C} Crocker, R.~M., \& Jones, D.~I., \& Aharonian, F. \& Law, C.~J.
	\& Melia, F., \& Oka, T., \& Ott, J. \ 2011, MNRAS, accepted



\bibitem[Cuadra et 
al.(2003)]{2003A&A...411..405C} Cuadra, J., Nayakshin, S., \& Sunyaev, R.\ 2003, \aap, 411, 405 

\bibitem[Cuadra et al.(2006)]{2006MNRAS.366..358C} Cuadra, J., Nayakshin, 
S., Springel, V., \& Di Matteo, T.\ 2006, \mnras, 366, 358 

\bibitem[Cuadra et al.(2008)]{2008MNRAS.383..458C} Cuadra, J., Nayakshin, 
S., \& Martins, F.\ 2008, \mnras, 383, 458 



\bibitem[Ferrarese et al.(2006)]{2006ApJ...644L..21F} Ferrarese, L., C{\^o}t{\'e}, P., Dalla Bont{\`a}, E., 
	Peng, E.~W., Merritt, D., Jord{\'a}n, A., Blakeslee, J.~P., 
	Ha{\c s}egan, M., Mei, S., Piatek, S.,Tonry, J.~L., 
	West, M.~J. \ 2006, \apj, 644, L21 



\bibitem[\protect\citeauthoryear{Genzel et al.}{2003}]{2003ApJ...594..812G} 
Genzel R., et al., 2003, ApJ, 594, 812 

\bibitem[Ghez et al.(2005)]{GhezEtal05} Ghez, A.~M., Salim, S., 
Hornstein, S.~D., Tanner, A., Lu, J.~R., Morris, M., Becklin, E.~E., 
\& Duch{\^e}ne, G.\ 2005, \apj, 620, 744 

\bibitem[Ghez et al.(2008)]{2008ApJ...689.1044G} Ghez, A.~M., et al.\ 2008, 
\apj, 689, 1044 




\bibitem[Goodman(2003)]{2003MNRAS.339..937G} Goodman, J.\ 2003, \mnras, 
339, 937 

\bibitem[Greene et al.(2010)]{2010ApJ...721...26G} Greene, J.~E., Peng, C.~Y., Kim, M., Kuo, C.~Y., 
	Braatz, J.~A., Violette Impellizzeri, C.~M., Condon, J.~J., 
	Lo, K.~Y., Henkel, C., Reid, M.~J.\ 2010, \apj, 721, 26 




\bibitem[Guo \& Mathews(2011)]{arXiv:1103.0055v1} Guo, F., \& Mathews, W. ~G.\ 2011, arXiv:1103.0055v1
339, 937 

	 

\bibitem[Hobbs 
\& Nayakshin(2009)]{2009MNRAS.394..191H} Hobbs, A., \& Nayakshin, S.\ 2009, \mnras, 394, 191 

\bibitem[Hobbs et al.(2010)]{2010arXiv1001.3883H} Hobbs, A., Nayakshin, S., 
Power, C., \& King, A.\ 2010, arXiv:1001.3883 


\bibitem[\protect\citeauthoryear{King}{2003}]{2003ApJ...596L..27K} King A.~R., 
2003, ApJ, 596, L27 

\bibitem[\protect\citeauthoryear{King}{2005}]{2005ApJ...635L.121K} King A.~R., 
2005, ApJ, 635, L121 

\bibitem[\protect\citeauthoryear{King}{2010}]{2010MNRAS.402.1516K} King 
A.~R., 2010a, MNRAS, 402, 1516 

\bibitem[\protect\citeauthoryear{King}{2010}]{2010MNRAS.408L..95K} King 
A.~R., 2010b, MNRAS, 408, L95 

\bibitem[\protect\citeauthoryear{King \&
    Pringle}{2007}]{2007MNRAS.377L..25K} King A.~R., Pringle J.~E.,
  2007, MNRAS, 377, L25


\bibitem[\protect\citeauthoryear{King \&
    Pounds}{2003}]{2003MNRAS.345..657K} King A.~R., Pounds K.~A.,
  2003, MNRAS, 345, 657

\bibitem{} King, A.~R., Zubovas, K., Power, C., \ 2011, arXiv:1103.1702, accepted by MNRAS

\bibitem[\protect\citeauthoryear{Kinney et al.}{2000}]{2000ApJ...537..152K} 
Kinney A.~L., Schmitt H.~R., Clarke C.~J., Pringle J.~E., Ulvestad J.~S., 
Antonucci R.~R.~J., 2000, ApJ, 537, 152 

\bibitem[Kolykhalov 
\& Syunyaev(1980)]{1980SvAL....6..357K} Kolykhalov, P.~I., \& Sunyaev, R.~A.\ 1980, Soviet Astronomy Letters, 6, 357 

\bibitem{}Kormendy, J., Bender, R., Cornell, M.E., 2011, {\it Nat} 469, 374

\bibitem[Levin \& Beloborodov(2003)]{2003ApJ...590L..33L} Levin, Y., \&
  Beloborodov, A.~M.\ 2003, \apjl, 590, L33


\bibitem[Lu et al.(2009)]{2009ApJ...690.1463L} Lu, J.~R., Ghez, A.~M., 
Hornstein, S.~D., Morris, M.~R., Becklin, E.~E., 
\& Matthews, K.\ 2009, \apj, 690, 1463 



\bibitem[\protect\citeauthoryear{McLaughlin, King, \&
    Nayakshin}{2006}]{2006ApJ...650L..37M} McLaughlin D.~E., King
  A.~R., Nayakshin S., 2006, ApJ, 650, L37



\bibitem[Mertsch \& Sarkar (2011)]{MS11} Mertsch, P. \& Sarkar, S.\ 2011, arXiv:1104.3585 










\bibitem[Morris 
\& Serabyn(1996)]{MS96} Morris, M., \& Serabyn, E.\ 1996, \araa, 34, 645 

\bibitem[Morris et al.(1999)]{MorrisEtal99} Morris, M., Ghez, A.~M., 
\& Becklin, E.~E.\ 1999, Advances in Space Research, 23, 959 

\bibitem[Muno et al.(2007)]{2007ApJ...656L..69M} Muno, M.~P., Baganoff, 
F.~K., Brandt, W.~N., Park, S., \& Morris, M.~R.\ 2007, \apjl, 656, L69 





\bibitem[Narayan(2002)]{2002luml.conf..405N} Narayan, R.\ 2002, Lighthouses 
of the Universe: The Most Luminous Celestial Objects and Their Use for 
Cosmology, 405 


\bibitem[\protect\citeauthoryear{Nagar \&
    Wilson}{1999}]{1999ApJ...516...97N} Nagar N.~M., Wilson A.~S.,
  1999, ApJ, 516, 97

\bibitem[Nayakshin \& Cuadra(2005)]{2005A&A...437..437N} Nayakshin, S., \&
  Cuadra, J.\ 2005, \aap, 437, 437

\bibitem[Nayakshin et al.(2006)]{2006MNRAS.366.1410N} Nayakshin, S., 
Dehnen, W., Cuadra, J., \& Genzel, R.\ 2006, \mnras, 366, 1410 

\bibitem[Nayakshin et al.(2007)]{2007MNRAS.379...21N} Nayakshin, S., 
Cuadra, J., \& Springel, V.\ 2007, \mnras, 379, 21 


\bibitem[\protect\citeauthoryear{Nayakshin, Wilkinson, \&
    King}{2009}]{2009MNRAS.398L..54N} Nayakshin S., Wilkinson M.~I.,
  King A., 2009, MNRAS, 398, L54

\bibitem[Paczynski(1978)]{1978AcA....28...91P} Paczynski, B.\ 1978, Acta
  Astronomica, 28, 91

\bibitem[Paumard et 
al.(2004)]{2004A&A...426...81P} Paumard, T., Maillard, J.-P., \& Morris, M.\ 2004, \aap, 426, 81 

\bibitem[Paumard et al.(2006)]{PaumardEtal06} Paumard, T., et al.\ 
2006, \apj, 643, 1011 

\bibitem[Ponti et al.(2010)]{2010ApJ...714..732P} Ponti, G., Terrier, R., 
Goldwurm, A., Belanger, G., \& Trap, G.\ 2010, \apj, 714, 732 


\bibitem[\protect\citeauthoryear{Pounds et al.}{2003}]{2003MNRAS.345..705P} 
Pounds K.~A., Reeves J.~N., King A.~R., Page K.~L., O'Brien P.~T., Turner 
M.~J.~L., 2003a, MNRAS, 345, 705 

\bibitem[\protect\citeauthoryear{Pounds et al.}{2003}]{2003MNRAS.346.1025P} 
Pounds K.~A., King A.~R., Page K.~L., O'Brien P.~T., 2003b, MNRAS, 346, 1025 

\bibitem[Revnivtsev et 
al.(2004)]{2004A&A...425L..49R} Revnivtsev, M.~G., et al.\ 2004, \aap, 425, L49 




\bibitem[Sch{\"o}del et al.(2002)]{SchoedelEtal02} Sch{\"o}del, R., et 
al.\ 2002, \nat, 419, 694 


\bibitem[Sch{\"o}del et 
al.(2009)]{2009A&A...502...91S} Sch{\"o}del, R., Merritt, D., \& Eckart, A.\ 2009, \aap, 502, 91 



\bibitem[\protect\citeauthoryear{Soltan}{1982}]{1982MNRAS.200..115S} Soltan 
A., 1982, MNRAS, 200, 115 



\bibitem[\protect\citeauthoryear{Shakura \&
    Sunyaev}{1973}]{1973A&A....24..337S} Shakura N.~I., Sunyaev R.~A.,
  1973, A\&A, 24, 337

\bibitem[\protect\citeauthoryear{Silk 
\& Rees}{1998}]{1998A&A...331L...1S} Silk J., Rees M.~J., 1998, A\&A, 331, L1 

\bibitem[Snowden et al.(1997)]{SnowdenEtal97} Snowden, S.~L., et al.\ 
1997, \apj, 485, 125 

\bibitem[Sofue(2000)]{Sofue00} Sofue, Y.\ 2000, \apj, 540, 224 

\bibitem[Sofue(2003)]{Sofue03} Sofue, Y.\ 2003, PASJ, 55, 445 

\bibitem[\protect\citeauthoryear{Su, Slatyer, \&
    Finkbeiner}{2010}]{2010ApJ...724.1044S} Su M., Slatyer T.~R.,
  Finkbeiner D.~P., 2010, ApJ, 724, 1044

\bibitem[Thompson et al.(2005)]{2005ApJ...630..167T} Thompson, T.~A., 
Quataert, E., \& Murray, N.\ 2005, \apj, 630, 167 

\bibitem[Tremaine et al.(2002)]{2002ApJ...574..740T} Tremaine, S., et al.\ 
2002, \apj, 574, 740 

\bibitem[Totani(2006)]{Totani06} Totani, T.\ 2006, PASJ, 58, 965



\bibitem[Wardle 
\& Yusef-Zadeh(2008)]{2008ApJ...683L..37W} Wardle, M., \& Yusef-Zadeh, F.\ 2008, \apjl, 683, L37 





\end{thebibliography}
\end{document}